\documentclass[epj]{svjour}
\usepackage{graphics}
\usepackage{color}
\begin{document}
\title{Existence and Switching Behavior \\
       of Bright and Dark Kerr Solitons \\
       in Whispering-Gallery Mode Resonators  \\
       with Zero Group-Velocity Dispersion}
\author{Jimmi H. Talla Mb\'e\inst{1,2,3}, Carles Mili\'an\inst{4} \and Yanne K. Chembo\inst{5,6}
}                     
%
%
\institute{Laboratory of Electronics, Automation and Signal Processing, Department of Physics, \\
           University of Dschang, P.O. Box 67, Dschang, Cameroon
\and       Laboratory of Modelling and Simulation in Engineering, Biomimetics and Prototypes, Department of Physics,\\
           University of Yaound\'e I,  P. O. Box 812, Yaound\'e, Cameroon
\and       African Center of Excellence for Information and Communication Technologies (CETIC), \\
           Polytechnic School of Yaound\'e, P.O. Box 8390, Yaound\'e, Cameroon
\and       Centre de Physique Th\'eorique, CNRS, Ecole Polytechnique, F-91128 Palaiseau, France
\and       Optics Department, FEMTO-ST Institute, CNRS \& Univ. Bourgogne-Franche-Comt\'e, \\
           15B Avenue des Montboucons, 25030 Besan\c con cedex, France
\and       GeorgiaTech-CNRS Joint International Laboratory [UMI 2958], Atlanta Mirror Lab, \\
           School of Electrical and Computer Engineering, 777 Atlantic Dr NW,
           Atlanta GA 30332-0250, USA.}

\date{Received: date / Revised version: date}
%
\abstract{We use the generalized Lugiato-Lefever model to investigate the phenomenon of
Kerr optical frequency comb generation when group-velocity dispersion is null.
In that case, the first dispersion term that plays a leading role is third-order dispersion.
We show that this term is sufficient to allow for the existence of both bright and dark solitons.
We identify the areas in the parameter space where both kind of solitons can be excited inside the resonator.
We also unveil a phenomenon of hysteretic switching between these two types of solitons when
the power of the pump laser is cyclically varied.
} 

\authorrunning{Talla Mb\'e, Mili\'an and Chembo}
\titlerunning{Solitons in WGM resonators with zero GVD}

\maketitle

\section{Introduction}

The topic of Kerr optical frequency comb using ultra-high $Q$ whispering gallery mode resonators has been the focus of extensive research in recent years~\cite{Review_Kerr_combs_Science,Nanophotonics}.
These combs are obtained after pumping these Kerr-nonlinear cavities with a resonant continuous-wave (CW) laser.
Above a certain threshold, the small volume of confinement, high photon density and long photon lifetime contribute to the excitation of the neighboring cavity eigenmodes  through four-wave mixing (FWM) interactions of the kind
$\hbar \omega_m + \hbar \omega_p \rightarrow \hbar \omega_n + \hbar \omega_q$, where two input photons $m$ and $p$ interact coherently via the Kerr nonlinearity to yield two output photons $n$ and $q$.
This FWM induces a global coupling between the modes, which potentially results in the excitation of up to several hundred spectral lines.
This cascade of photonic interactions yields the so-called Kerr optical frequency comb, which is a set of equidistant spectral components in the Fourier domain.
Many features of the nonlinear and quantum properties of these combs have already been analyzed in depth in several research works
~\cite{YanneNanPRL,YanneNanPRA,Matsko_OL_2,PRA_Yanne-Curtis,Coen,OL_phaselocking,Chaos_paper,PRA_Rogue_wave,PRA_Unified,PRA_Quantum}.
From the applications standpoint, Kerr combs have been found to be of particular relevance in many areas, including aerospace and telecommunication engineering, spectroscopy, and microwave/lightwave frequency
synthesis~\cite{DelhayeKipp,Lipson_NatPhot,Nature_Ferdous,PRL_Vahala_2012,PRL_NIST,PfeifleNatPhot,NIST_Optica,Self_injection_NIST,Kipp_Soliton,PRL_WDM,SR_Huang,Saleh2016OE,Saleh2016PJ}.

It is noteworthy that Kerr comb generation requires the frequency shift generated by self-phase modulation (SPM) to be compensated by both the laser frequency detuning and the overall (chromatic and geometric) dispersion of the resonator.
From this requirement, one can foreshadow the central role that dispersion plays in the Kerr comb generation process.
In general, studies in Kerr optical frequency comb generation only focus on the two distinct signs for the second-order dispersion parameter (also known as group-velocity dispersion, or GVD), generally referred to as normal (positive GVD) and anomalous (negative GVD) dispersion regimes.
Kerr comb generation in both regimes has been the focus of a large amount of research work, that has allowed to identify the various possible solutions arising in each case, and which include rolls (also referred to as Turing patterns), spatiotemporal chaos, and solitons of various forms: bright/dark, breathers, and molecules~\cite{PRA_Unified,Parra_Rivas}.
Bright solitons are usually found in the anomalous dispersion regime~\cite{PRA_Unified,Kipp_Soliton,Parra_Rivas},
in contrast to dark solitons that are generally obtained in the normal dispersion
regime~\cite{PRA_Unified,Liang,Huang_PRL,Xue,Lobanov,Parra_Rivas_OL1,Parra_Rivas_PRA,Parra_Rivas_arXiv}.

However, the case of zero GVD has not been analyzed in depth, despite its relevance from a bifurcation analysis point of view.
On the one hand, it is well known that very small GVD is desirable in order to obtain broadband combs, and therefore it is important to analyze what occurs when second-order dispersion is strictly null, as this permits to understand how the system behaves asymptotically on each side of the zero-GVD limit~\cite{Milian_OE}.
On the other hand, when the second-order dispersion term  is null, the higher orders of dispersion should be considered and in this case the third-order dispersion (or TOD) plays the leading role.
In both cases, the spectro- and spatio-temporal properties of the comb display some singular features that are still to be investigated and understood in depth.

{
By setting the pump exactly at the zero GVD, where $b_3>0$, all excited comb lines fall in the normal (anomalous) if blue (red)-shifted from the pump. Any soliton (bright or dark) formed around the pump is strongly affected by the recoil effect, associated to the resonant radiation tail, which helps locating the soliton core firmly in the normal or anomalous GVD for bright or dark solitons, respectively (see ref.~\cite{Milian_OE} for the case of a bright soliton in the anomalous GVD and the pump under the normal GVD). Therefore, under this conditions, coexistence of the bright and dark solitons is highly expected, and indeed possible. In fact, there is no reason for both types of solitons to coexist in both the normal and anomalous GVD regime for the pump.
}

\begin{figure}
\begin{center}
\resizebox{0.4\textwidth}{!}{%
  \includegraphics{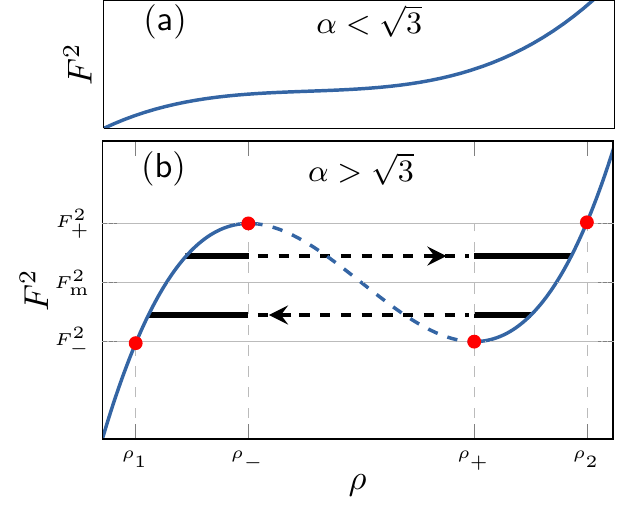}
}
\end{center}
\caption{(Color online) Laser pump versus the stable equilibrium point of the intra-cavity energy. The analytical values are plotted from Eq.~(\ref{Equilibria}). For (a), $\alpha < \sqrt{3}$: only one stable equilibrium. For (b), $\alpha > \sqrt{3}$: two stable (solid curve) and one unstable (dotted curve) equilibrium points. The attractive curve $\rho_{\rm d}$ represents all the values of $\rho$ lower than $\rho_{-}$ whereas, the attractive curve $\rho_{\rm u}$ represents all the values of $\rho$ greater than $\rho_{+}$. The solid parts of the arrows is used to emphasize on the fact that one among $\rho_{\rm u}$ and $\rho_{\rm d}$ can be more attractive than other (the solid part is longer in the side of attractive line), becoming therefore the final predominant amplitude of the flat state.  The dash part of an arrow emphasizes on the unstable equilibrium $\rho_{\rm i}$. The axis ticks have been removed for a general interpretation. But, in (a), $\alpha = 1.5$ while for (b) $\alpha = 3.5$.}
\label{Plotting_Equilibrium_Stability_2}       
\end{figure}

In this article, we aim to address this problem, and analyze the dynamical properties
of soliton Kerr combs when GVD is null.
The structure of the article is therefore the following.
In the next section, we present the model under investigation, which is a generalized Lugiato-Lefever equation.
Section~\ref{zero_disp} is devoted to the case of null overall dispersion, where only flat states are allowed.
We then analyze in the Sec.~\ref{ZeroGVD} the effect of third-order dispersion, which is the first dispersion term to be
accounted for when GVD is null.
The last section concludes the article.

\section{Model}
\label{Model}

Kerr comb generation is usually investigated using the mean-field Lugiato-Lefever equation (LLE)~\cite{LL},
which was found to provide an accurate insight into the intracavity spatiotemporal dynamics of the system.
The generalized LLE for the study of Kerr comb generation can be explicitly written in a dimensionless form as
\begin{equation}
\frac{\partial \psi}{\partial t} = F-(1+i\alpha) \psi + i |\psi|^{2} \psi
                                    +i \sum_{n= 2}^{n_{\rm max}} i^n \frac{b_n}{n!}
                                                                 \frac{\partial^{n}\psi}{\partial \theta^{n}} \, ,
\label{DispersiveLLE}
\end{equation}
where the variable $\psi(\theta , t)$ stands for the complex slowly-varying envelope of the total intracavity field.
The dimensionless time $t$ is scaled with regards to $2 \tau_{\rm ph}$, where $\tau_{\rm ph}=1/\omega_{\rm tot}$ is the photon lifetime of the loaded cavity and $\Delta \omega_{\rm tot}$ is the total (or loaded) linewidth of the pumped resonance.
The variable $\theta \in [-\pi,\pi]$ stands for the azimuthal angle along the circumference of the resonator.

\begin{figure*}
\begin{center}
\resizebox{0.7\textwidth}{!}{%
  \includegraphics{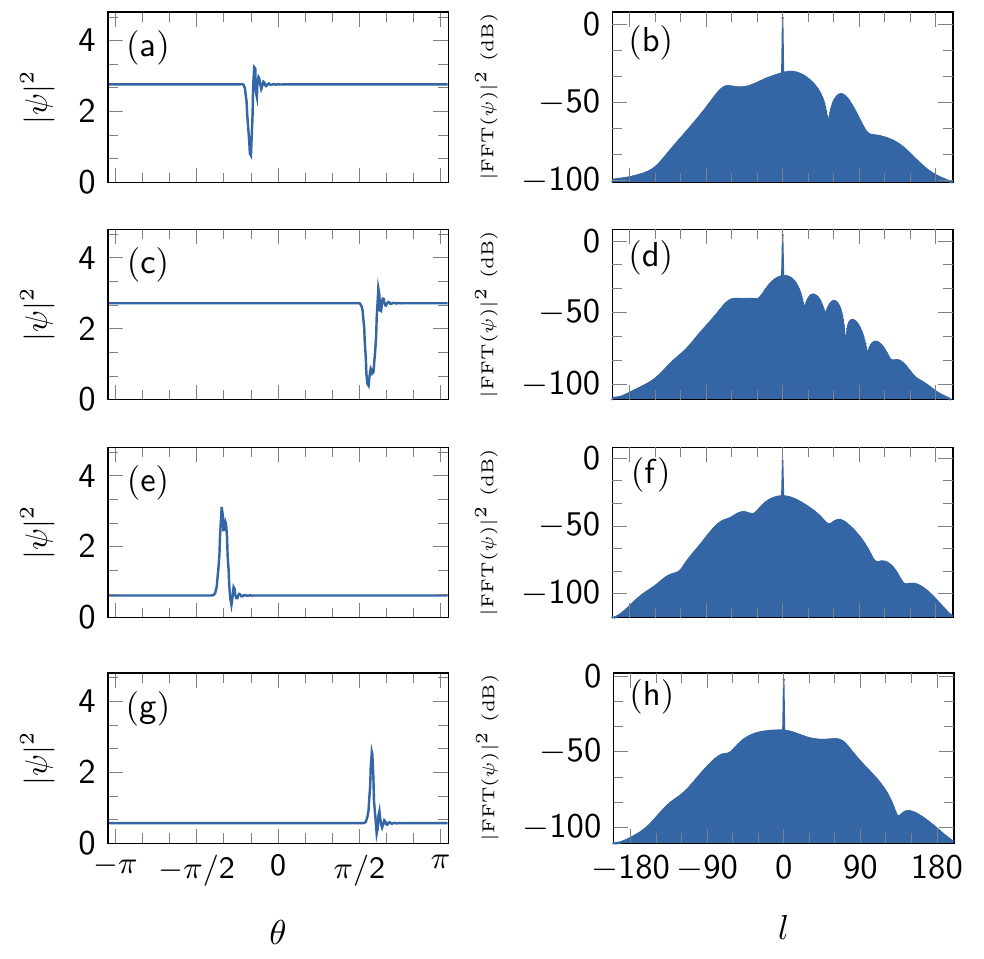}
}
\end{center}

\caption{(Color online) {Soliton spatial waveforms (left column) and the corresponding} frequency combs (right column) for the pump around $F_{\rm m}=1.68$.
As the pump is monitored from $1.72$ [(a) and (b)] to $1.64$ [(g) and (h)],
the soliton waveform changes as well from for both dark and bright cavity solitons.
The frequency combs have a single-FSR spacing, but their amplitudes undergo some changes [(b), (d), (f), (h)].
For these plots, $b_{3}=4.08 \times 10^{-5}$, $\alpha =2.5$, [(a) and (b)] $F=1.72$,  [(c) and (d)]  $F=1.69$,  [(e) and (f)] $F=F_{m}$ and  [(g) and (h)] $F=1.64$.}
\label{Soliton_Type_And_Combs}       
\end{figure*}

The dimensionless LLE has periodic boundary conditions in the angular variable and is characterized by the parameters $F$ (pump), $\alpha$ (frequency detuning), and $b_n$ (dispersion parameters at the $n$-th order, with $2 \leq n \leq n_{\rm max}$).
More specifically, the dimensionless and real-valued pump field $F$ is related to the pump power $P$ (in watts) as
$F = [{8 g_0 \Delta \omega_{\rm ext}}/{\Delta \omega_{\rm tot}^3}]^{\frac{1}{2}} [{P}/{\hbar \omega_{\rm las} } ]^{\frac{1}{2}}$,
where 
$\omega_{\rm las} $ is the laser angular frequency, $g_0 = n_2 c \hbar \omega_{\rm las}^2/n_0^2 V_0$ is the nonlinear gain with
$n_0$ and $n_2$ being respectively the linear and nonlinear refraction indices of the bulk material, $V_0$ is the effective volume of the pumped mode, $c$ is the velocity of light in vacuum, $\hbar$ is the reduced Planck constant, while $\Delta \omega_{\rm ext}$ is the external (or coupling) linewidth of the pumped resonance.
The detuning parameter is $\alpha = - {2 \sigma }/{\Delta \omega_{\rm tot}}$, where
$\sigma = \omega_{\rm las} -\omega_{\rm res}$ is the difference between
the pump laser and the cold-cavity resonance angular frequencies.
Finally, $b_n$ stands for the $n$-th order overall dispersion parameter of the resonator.
It is also important to note that the intracavity power in watts can be simply recovered as
$|{E}|^2 = [\hbar \omega_{\rm las} / 2 g_0 \tau_{\rm ph} T_{_{\rm FSR}} ] |\psi|^2$, where $T_{_{\rm FSR}}= 2\pi/ \Omega_{_{\rm FSR}}$ is the intracavity round-trip time.
For crystalline disk-resonators, the photon lifetime $\tau_{\rm ph}$ can be longer than $1$~$\mu$s
(see refs.~\cite{Jove,OL_BaF2,OL_SrF2,Lin_OExp}) while it is of the order of few ns for integrated resonators.
The GVD and TOD respectively correspond to $b_2$ and $b_3$.
By convention, the anomalous GVD regime is defined by $b_2<0$ while normal GVD corresponds to $b_2>0$.

It is known that the LLE can describe the dynamics of the intracavity field with outstanding precision.
For example, in ref.~\cite{Chaos_paper}, the experimental spectra involved more than $300$~modes,
and where shown to agree with the theoretical one over a dynamical range greater than $80$~dB.
We will investigate in the next sections these dispersion parameters on the spatio-temporal dynamics
of the intracavity field.

\begin{figure}
\begin{center}
\resizebox{0.35\textwidth}{!}{%
  \includegraphics{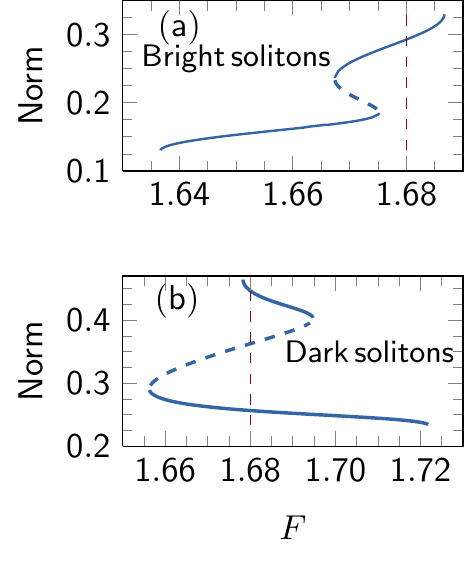}
}
\end{center}
\caption{(Color online) Soliton norm branches as a function of the pump field for (a) bright and (b) dark solitons when the pump wavelength is exactly set at the zero GVD value ($b_2=0$). The only higher-order dispersion term accounted for is TOD with $b_3=4.08\times10^{-5}$ and $\alpha=2.5$. Solid (dashed) curves indicate stable (unstable) branches.
The vertical dashed lines marks $F=F_{\rm m}=1.68$.}
\label{Bright_Dark_Solitons_Stability}       
\end{figure}

\section{Dynamics of the system with null overall dispersion}
\label{zero_disp}

The dynamics of the resonator when the overall dispersion is null is obtained
by uniformly setting the dispersion terms $b_n$ to zero in Eq.~(\ref{DispersiveLLE}).
The equilibria $\psi_{\rm e}$ are obtained by setting the temporal derivative to zero as well,
and they are solutions of the following nonlinear algebraic equation:
\begin{equation}\label{Equilibria}
  F^{2}=\rho^{3}-2\alpha \rho^{2}+(\alpha^{2}+1)\rho \equiv G(\alpha,\rho),
\end{equation}
with $\rho=|\psi_{\rm e}|^{2}$ being the intracavity power at equilibrium.
Because they are independent of time and space, the equilibria $\psi_{\rm e}$ are homogenous steady state solutions which are represented by a single spectral line in the Fourier domain.
The extrema of the function $G(\alpha,\rho)$ are
\begin{eqnarray}\label{def_extrema}
F_{\pm}^{2} & =& G[\alpha,\rho_{\mp}(\alpha)] \nonumber \\
            & =& \frac{2}{27} [\alpha (\alpha^{2}+9)\pm \sqrt{(\alpha^{2}-3)^{3}}]
\end{eqnarray}
with $\rho_{\mp}(\alpha)=[2 \alpha \mp \sqrt{\alpha^{2}-3}]/3$.
For a given pump intensity $F^{2}$, Eq.~(\ref{Equilibria}) shows that we have only one equilibrium if $\alpha < \sqrt{3}$, but up to three equilibria
$\rho_{\rm d}< \rho_{\rm i}<\rho_{\rm u}$ when $\alpha > \sqrt{3}$, with the subscripts d, i and u standing for down, intermediate and up, respectively.
It is well known that the down- and uppermost solutions are stable, while the intermediate one is unstable (see, e.g., ref.~\cite{BarashPRE} for a thorough analysis of the flat state solutions).

In order to analyze the stability of these equilibria, it is convenient to explicitly split $\psi$ into its real and imaginary parts so that Eq.~(\ref{DispersiveLLE}) can be rewritten under the form of a two-dimensional flow with real variables.
The eigenvalues of the corresponding Jacobian matrix around the flat-states are given by:
\begin{eqnarray}\label{eigenvalues}
    \lambda_{\pm} &=& -1 \pm \sqrt{1-\frac{\partial G}{\partial \rho}} \nonumber \\
                  &=& -1 \pm \sqrt{(3\rho-\alpha)(\alpha-\rho)}
\end{eqnarray}
The analysis of these eigenvalues reveals that no local bifurcation can occur in the system except
when $\rho=\rho_\pm$ since, as one of the eigenvalues is laying on the imaginary axis ($\lambda_+=0$) and the second has a negative real part ($\lambda_- < 0$). The equilibria are therefore non-hyperbolic in this case.

For a pump field such that  $F_{-}^2 <F^2 < F_{+}^2$, the system is driven in the hysteresis area.
The asymptotic flat state will emerge from a competition between $\rho_{\rm d}$ and $\rho_{\rm u}$, which are both stable attractors. In fact, the final state will critically depend on the value $F^2$ of the pump with regards to the median pump power:
\begin{equation}\label{Middle_Pump}
   F_{\rm m}^2=\frac{1}{2} \, (F_{-}^2+F_{+}^2)= \frac{2\alpha(\alpha^{2}+9)}{27}\, .
\end{equation}
For a pump $F < F_{\rm m}$, the amplitude of the flat state will be $\rho_{\rm d}$,
while for  $F > F_{\rm m}$, the system will converge to $\rho_{\rm u}$.
This phenomenology, which is described in Fig.~\ref{Plotting_Equilibrium_Stability_2}b, predicts that perturbations to the singular problem of null dispersion will strongly depend on the value of the pump $F^2$ with regards to the median power $F_{\rm m}^2$.

In the next section, we will investigate the effect of TOD when GVD is still null, and analyze the deviation form of 
the singular manifolds in the spatiotemporal domain.

\section{Bright and dark solitons in the presence of third-order dispersion}
\label{ZeroGVD}

We now consider Eq.~(\ref{DispersiveLLE}) when all the dispersion coefficients are equal to zero, except the TOD coefficient $b_3$.

The effect of TOD on dissipative structures has already been analyzed in several research works.
Some of these investigations were focused on the reversibility breaking in the system which is responsible for the
soliton drift, while other works analyzed specific solutions such as time-varying
solitons~\cite{Milian_OE,Tlidi1,Tlidi2,Taki,Tlidi3,Rivas,Leo_PLA}.
It is important to note that owing to geometrical dispersion, the overall dispersion of a resonator
can in some cases be engineered via its shape to yield a dispersion profile closely corresponding to arbitrary
configurations~\cite{OE_Grudinin,OL_zhang,STQE_Zhang,OL_Wang,PRA_Bao,Grudinin_Optica}.

\begin{figure*}[t]
\begin{center}
\resizebox{1.0\textwidth}{!}{%
  \includegraphics{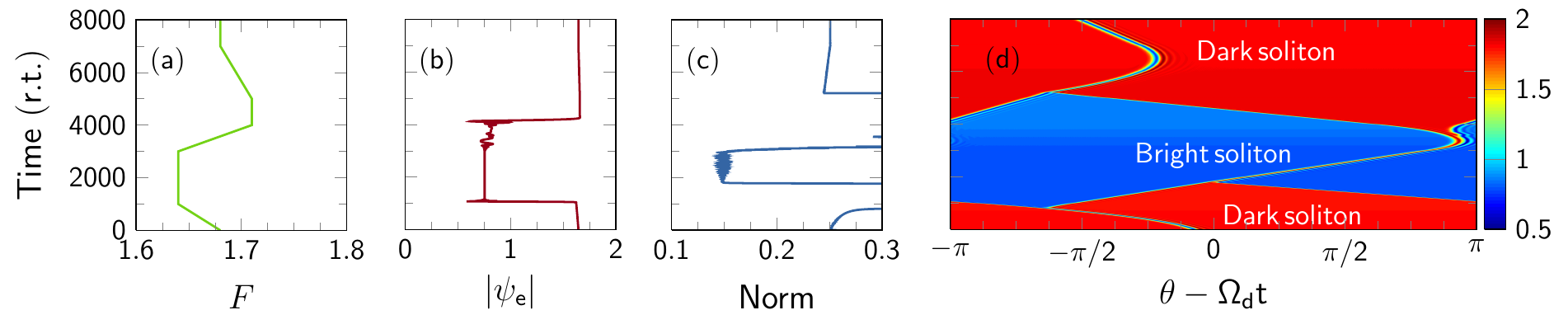}
}
\end{center}

\caption{(Color online) {Soliton switching amongst various branches when the pump power is cyclically varied.
The vertical axis of all figures is time in units of round-trips (r.t).
(a) Variations of the pump field $F$.
(b) Variations of the flat state $\psi_{\rm e}$.
(c) Variations of the norm $\mathsf{N}$.
(d) Color-coded intracavity field intensity, showing the various soliton branches (once the drift $\Omega_{\rm d}$ is removed).}}
\label{Multiplot_switching}       
\end{figure*}

When TOD is accounted for, non-trivial states may arise in the system close to, or inside the hysteresis area.
In particular, for a pump selected at a particular median value \textbf{$F_{\rm m}=1.68$} or just around it,
the system preferably converges towards a cavity soliton instead of flat states.
Bright and dark solitons yielding Kerr combs with single-FSR spacing are obtained.
The spatiotemporal profiles and the corresponding spectra are displayed in
Fig.~\ref{Soliton_Type_And_Combs}.
Despite the fact that the type of soliton (bright or dark) depends on the initial condition, bright cavity solitons are preferably obtained when the pump is below $F_{\rm m}$, whereas dark cavity solitons preferably emerge in the opposite case. For pump values far away from $F_{\rm m}$, the system asymptotically converges to a flat state.

Some spatial and spectral profiles of solitons are displayed in Fig.~\ref{Soliton_Type_And_Combs}.
Previous works demonstrated that large values of the TOD had the effect of stabilizing the frequency
combs~\cite{Milian_OE,Rivas,wang2015} even in the presence of Raman scattering~\cite{Milian_PRA}.

In order to investigate the  branches of stable solitons that can emerge in the system, it is
convenient to define a norm that can unambiguously discriminate a soliton from the trivial flat state.
There are indeed various ways to define such a norm, but one of the simplest is the following:
\begin{equation}
\mathsf{N} \equiv\int_{-\pi}^{+\pi}\left|\psi_{\rm s}-\psi_{\rm e}\right| \, d\theta \, ,
\end{equation}
where $\psi_{\rm s}(\theta)$ is the soliton spatial profile along the resonator and $\psi_{\rm e}$ are the equilibria of Eq.~(\ref{DispersiveLLE}).
Figure~\ref{Bright_Dark_Solitons_Stability} shows the variation of this soliton norm $\mathsf{N}$ as a function of the pump field $F$, for bright and dark solitons with $b_3>0$.

Figure~\ref{Bright_Dark_Solitons_Stability} also indicates that the solitons can become unstable via an hysteretic
path. As a consequence, we can excite the solitons of the same type (bright or dark) with two different amplitudes depending on initial conditions.
It can also be seen that bright and dark solitons can coexist depending on the initial condition.
{As suggested in refs.~\cite{Parra_Rivas_OL1,Parra_Rivas_PRA,Parra_Rivas_arXiv}, the appropriate
way to understand why both bright and dark solitons can coexist consists in analyzing the switching waves connecting the homogeneous steady state solutions $\rho_{\rm d}$ and $\rho_{\rm u}$, where $\rho_{\rm d}$ and $\rho_{\rm u}$ (with $\rho = |\psi_{\rm e}|^2$) are the stable flat states.
The coexistence of bright and dark solitons was already demonstrated in the regime of normal GVD, and attributed to a joint contribution of the third-order dispersion and large detuning~\cite{Parra_Rivas_arXiv}.

{
We note here that the coexistence of stable bright and dark solitons brings interesting dynamical features to the system.
This is illustrated in Fig.~\ref{Multiplot_switching}(a), where the pump, $F$ is changed with time.
The initial state of the resonator is given by the dark soliton at $F=1.68$ and norm $\simeq 0.25$
(see branch in Fig. \ref{Multiplot_switching}b). As time goes on, $F$ is decreased and the initial
dark soliton now follows a path in a branch (a signature of this phenomenology is the soliton velocity change with $F$) until the pump falls outside the existence limit defined by this branch.
At this point, because bright solitons do exist for when dark solitons do not, the dark pulse bifurcates to a bright soliton
of norm $\simeq 0.15$. At this point, $F$ is increased and swept through the bright soliton branch until it crosses the existence limit for bright solitons.
Here again, because dark solitons can still exist when bright ones do not, the bright pulse bifurcates back to the dark soliton found for $F=1.71$ and norm $\simeq 0.22$. 
The pump is decreased until the initial value of $F=1.68$, thereby closing the loop. 
}

{
At the dark-to-bright (bright-to-dark) soliton transitions, the background field switches from the upper to the lower (lower to upper) states of the WGM resonator Fig.~\ref{Multiplot_switching}(b). As shown in Fig.~\ref{Multiplot_switching}(b), an interesting feature of this effects is that background switching occurs within a significantly narrower range for the pump powers, $F$, than that defining the bistability loop of the flat states. Additionally, while the pump range to achieve switching is small, the intensity difference between upper and lower states is kept high, i.e., the bistability loop may be made narrow in $F$ but it comes at the price of diminishing the contrast between upper and lower states. 
Figure~\ref{Multiplot_switching}(c) shows how the norm changes with the cyclic pump variation, while 
Fig.~\ref{Multiplot_switching}(d) displays the intracavity field.
The color-coded representation of this field enables to identify unambiguously the various soliton branches that can be excited in the system.
}

We can gain even better understanding of the phenomena presented in Fig.~\ref{Multiplot_switching} by plotting 
the variation of the background flat state $|\psi_{\rm e}|$ as a function of the cyclically varying pump field $F$.
This representation is displayed in Fig.~\ref{hysteresis_cycle}, where the hysteretic cycle can be explicitly
identified.

\begin{figure}[b]
\begin{center}
\resizebox{0.45\textwidth}{!}{%
  \includegraphics{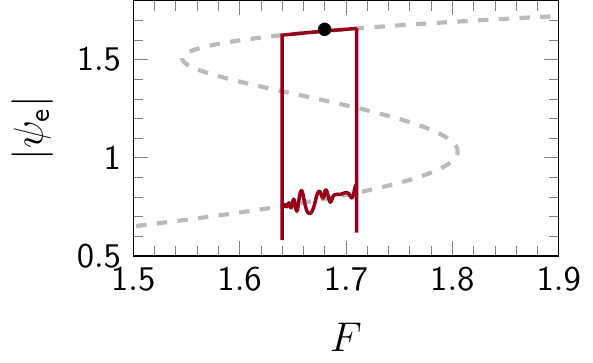}
}
\end{center}
\caption{(Color online) {Hysteretic cycle (in red) completed by $|\psi_{\rm e}|$ as $F$ is periodically ramped up and down. The temporal dynamics of both variables is exactly the one displayed in Figs.~\ref{Multiplot_switching}(a) and~(b). In the figure, the cycle begins on the black dot and goes counter-clockwise. 
The gray dashed line corresponds to the relationship defined by Eq.~(\ref{Equilibria}) between the pump field and the flat states.}}
\label{hysteresis_cycle}       
\end{figure}

\section{Conclusion}
\label{conclusion}

In this work, we have investigated the dynamics of Kerr optical frequency combs when the group-velocity dispersion is set to zero.
We have demonstrated that third-order dispersion is sufficient to lead to the emergence of
both bright and dark solitons.
These solitons have been shown to coexist in some range around an optimal value of the pump power that has been analytically defined. We have also analyzed the hysteretic switching between both types of solitons when the pump power is cyclically varied. 
Future work will focus on the consideration of other effects such as Raman scattering or pulsed-pumping
schemes~\cite{BarashPRE,Luo2015,PRA_Raman,OE_Bao,OE_Universal,OE_Raman_Lin} when the system is still very close or at zero group velocity dispersion. This research will enable the community to understand better how the Kerr comb span can be expanded to its maximum extent with the smallest energy footprint possible.


\begin{acknowledgement}
J. H. T. M. acknowledges financial support the
African-German Network of Excellence in Science (AGNES). Y. K. C. would like to acknowledge financial
support from the European Research Council (ERC)
through the project NextPhase, from the Centre National
d'Etudes Spatiales (CNES) through the project SHYRO,
from the R\'egion de Franche-Comt\'e through the project
CORPS, and from the Labex ACTION.
\end{acknowledgement}

%
%
%

%

\end{document}